# Viewpoint

## Fluctuation diamagnetism in high-temperature superconductors

**Steven A. Kivelson**
*Department of Physics, McCullough Bldg., 476 Lomita Mall, Stanford University, Stanford, CA 94305-4045, USA*

**Eduardo H. Fradkin**
*Department of Physics, University of Illinois, 1110 W. Green St., Urbana, IL 61801-3080, USA*

*The observation of a strong nonlinear diamagnetic signal well above $T_c$ in several cuprate superconductors constitutes clear evidence of the persistence of local superconducting correlations.*

Subject Areas: **Superconductivity**

---

**A Viewpoint on:**
**Diamagnetism and Cooper pairing above $T_c$ in cuprates**
Lu Li, Yayu Wang, Seiki Komiya, Shimpei Ono, Yoichi Ando, G. D. Gu and N. P. Ong
*Phys. Rev. B* **81**, 054510 (2010) – Published February 16, 2010

---

More than two decades after the discovery of the cuprate high-temperature superconductors, there is still healthy debate concerning the character of the highly anomalous "normal" state above the superconducting transition temperature $T_c$. An unresolved issue is the extent of the temperature and magnetic-field regime in which superconducting fluctuations persist, and how best to characterize this regime theoretically. One reason for this lack of clarity is that other types of order parameter fluctuations are also significant, making disentangling the contributions from the different kinds of fluctuations highly complex.

STM and ARPES measurements show evidence of incipient order above $T_c$, and of a single-particle gap that persists into the normal state. However, distinguishing a superconducting gap from a density-wave gap is a subtle task. An extremely revealing set of measurements has also been carried out [1, 2] on the Nernst effect, a thermoelectric phenomenon in which a voltage transverse to a temperature gradient is created in a conducting sample subjected to a magnetic field. The Nernst effect, however, is highly sensitive not only to superconducting fluctuations, but also to any type of order that leads to a reconstruction of the Fermi surface. *A priori*, the ideal method to detect short-range superconducting order would be to measure the magnetization. The diamagnetic response of an ordered superconductor is many orders of magnitude stronger than that of any other known state of matter, meaning that even when the superconducting correlation length is finite, the fluctuation diamagnetism can be large compared to the "background." Moreover, magnetization is a thermodynamic quantity, and as such is not subject to the uncertainties in interpretation that arise concerning dynamical and/or nonequilibrium properties.

In a paper by Lu Li and co-workers appearing in *Physical Review B*[3], the group of N. Phuan Ong in Princeton, US, together with collaborators at Tsinghua University, China, Central Research Institute of Electric Power Industry and Osaka University, both in Japan, and Brookhaven National Laboratory, US, have compiled the results of a major experimental study of the magnetization of several important families of cuprate high-temperature superconductors over a broad range of temperatures and magnetic fields. From these, they infer a field-dependent onset temperature, $T_M$, below which superconducting fluctuations are strong, as shown qualitatively in Fig.1. It is refreshing to find important results presented in a comprehensive and scholarly paper of adequate length. In particular, this paper takes care to justify the analysis of the raw data that is performed prior to its being interpreted.

The magnetization is measured using torque magnetometry, a technique that has the added benefit that there is essentially no spin contribution to the measured magnetization [4]. Li *et al.* observe a weakly-temperature-dependent paramagnetic signal at high temperatures ($T > T_M$), which they attribute to van Vleck paramagnetism, unrelated to superconducting fluctuations. Specifically, for $T > T_M$, the magnetization is linear in $H$ over the entire accessible range, and can be well fit by the expression $M = (A + BT)H$, with $A \gg BT$. Li *et al.* assume that this contribution is present, more or less unchanged, over the entire range of $T$ and $H$, and thus in their analysis they plot a "diamagnetic contribution" to the magnetization:

$$M_d(H,T) = M_{\exp}(T,H) - (A + BT)H, \qquad (1)$$

where $M_{\exp}$ is the measured magnetization. This subtraction is the only way in which the raw data is modified. None of the strong nonlinear $T$ or $H$ dependences observed by Li *et al.* can be artifacts of this subtraction. Although in much of the phase diagram the subtracted term is small compared to $M_{exp}$, close to $T = T_M$, where the diamagnetic fluctuations are becoming undetectably small, the subtraction has a large effect on the analysis. However, other than small uncertainties in the precise value of $T_M$, it seems unlikely that the subtraction could produce misleading results.

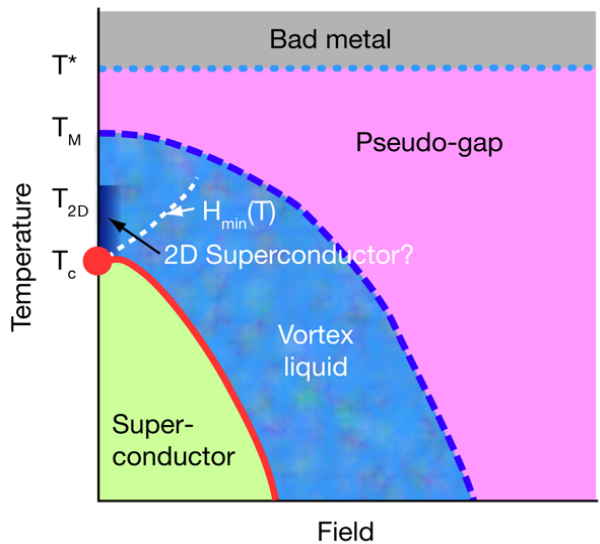

FIG. 1: Schematic phase diagram of a cuprate superconductor plotted as a function of the applied field ($H$) and temperature ($T$). The field-dependent crossover temperature $T_M$ bounding the vortex liquid region is shown as a green dashed line. The superconducting phase is bounded by the solid red line. The region in which evidence of unusual, possibly 2D superconducting behavior is found in some cuprates is indicated by the dark blue region above $T_c$. $T^*$ denotes the crossover between the "bad metal" phase—so called because in this state conventional transport theory breaks down—and the pseudogap phase, which recent experiments suggest may be related to the onset of electron-nematic order [2, 11]. (Illustration: Alan Stonebraker)

The boundary between the superconducting and normal states (the solid red line in Fig.1) marks a thermodynamic phase transition and is unambiguous in that the resistance is zero in the superconducting state and nonzero beyond it. The remaining lines in the phase diagram are, to the best of our current knowledge, crossovers from one state to another and hence less sharply defined. The region labeled "vortex liquid" is identified as a region of strong superconducting fluctuations because of several characteristic features of the magnetization curves: the magnetization is opposite in direction to $H$ (diamagnetic), it is large compared to (for example) the Landau diamagnetism in conventional metals, and it is a nonlinear function of $H$. $T_M$ is thus identified as the point at which $M_d$ vanishes, and where $M_{exp}$ changes from being a linear function of $H$ (for $T > T_M$) to nonlinear (for $T < T_M$).

The nonlinearity is an expected feature of superconducting fluctuations. For small $H$, one expects $M$ to grow linearly $M_d = -\chi_{fluc}H$, where $\chi_{fluc}$ is the diamagnetic susceptibility. However, for $H$ in excess of a characteristic field (that one might think of as being the mean-field $H_{c2}$), the magnetization should tend to zero. This leads one to expect that for temperatures between $T_M$ and $T_c$, $M$ should exhibit a minimum as a function of applied field at a nonzero, $H_{min}$. The fact that such behavior is observed in the "vortex liquid" region above $T_c$ is the strongest piece of evidence for a superconducting origin of the observed effects. The authors observed that in all the materials studied here, other than optimally doped YBa$_2$Cu$_3$O$_7$, $H_{min} \to 0$ as $T \to T_c$, but $H_{min}$ grows increasingly large with increasing $T$ above $T_c$, as shown schematically by the green dashed line in Fig.1. The large magnitude of $H_{min}$ at elevated temperatures implies that high applied fields (up to 45 T in the present paper) are needed to observe this field dependence clearly. Indeed, because the mean-field $H_{c2}$ in the high-temperature superconductors is so large, in most cases even experiments carried out up to 45 T only access fields somewhat in excess of $H_{min}$. The values of $H_{c2}$ at most temperatures less than $T_M$ are higher than have been achieved, so the high field portions of the schematic phase diagram in Fig.1 must be obtained by extrapolation.

Li *et al.* present another argument to confirm the superconducting origin of the observed diamagnetism. There is general consensus that the resistive transition at temperatures $T < T_c$ marks the boundary between a superconducting phase and a vortex liquid. The vortex liquid is characterized by strong, local, superconducting correlations and a well-defined short-distance superfluid stiffness that is degraded, at long distances, by vortex motion. It is therefore significant that the thermal evolution of the magnetization at moderate fields in the range $H_{min} < H < H_{c2}$, is smooth through $T_c$, and only shows a qualitative change as $T$ approaches $T_M$. Taken together, these various observations constitute extremely compelling evidence of a broad range of superconducting fluctuations above $T_c$ as a generic feature of the phase diagram of the cuprate high-temperature superconductors.

Two dramatic features of the data do not follow straightforwardly from simple considerations of superconducting fluctuations.

(1) At least in two families of cuprates, Bi$_2$Sr$_2$CaCu$_2$O$_{8+x}$ and Bi$_2$Sr$_{2-y}$La$_y$CuO$_6$, the low field magnetization exhibits apparently nonanalytic behavior, in that $M \sim |H|^x$ with $x < 1$ as $H \to 0$, for a range of temperature $T_{2D} > T > T_c$ (shown as the shaded bar in Fig.1). If this behavior truly extends to arbitrarily small $H$, it implies a divergent susceptibility, which in turn suggests that $T_{2D}$ is a critical point, suggesting a new, distinct, critical phase of matter, the onset of which occurs below this temperature. An idea of what such a phase might look like was proposed by Oganesyan *et al.*[5]. They observed that in a layered superconductor with *zero* Josephson coupling between planes, the critical phase below the Kosterlitz-Thouless transition temperature of the individual layers has a diamagnetic magnetization at small $H$ given by the scaling relation

$$|M| \sim (T_{2D} - T)\ln[H_c/|H|]. \quad (2)$$

While this is not quite of the same form as reported by Li *et al.* and in an earlier study by Wang *et al.*[6], it resembles the data at least to the extent that it produces a divergent susceptibility for a range of temperatures. An apparent problem with this interpretation is that even weak but finite interplane Josephson coupling inevitably leads to a 3D superconducting transition with $T_c > T_{2D}$. (A different interpretation of the vortex liquid as an "incompressible superfluid" was proposed by Anderson [7].)

(2) It is not obvious why $H_{\text{min}}$ should tend to vanish as $T \to T_c$. Given that it does, simple scaling arguments imply that near $T_c$, the characteristic field strength $H_{\text{min}} \sim \phi_o/\xi^2$ with $\phi_0 = hc/2e$.

It is not possible to extract accurate critical exponents from the published data; the existing data is consistent with the scaling behavior $H_{\text{min}} \sim (T - T_c)^{2\nu}$, with $2\nu \sim 1$. This is reasonable for a 3D classical critical point, or even a 2D Ising transition, but not remotely consistent with 3D $XY$ (Kosterliz-Thouless) scaling. Given the small magnitude of the interplane couplings in these materials, it would be surprising to find 3D critical scaling over a broad range of temperatures above $T_c$.

What are the implications of these two observations? To begin with, it is worth noting that a phase diagram with a similar topology to that shown in Fig. 1 was recently inferred [8] from transport and other measurements on the stripe-ordered high-temperature superconductor La$_{2-x}$Ba$_x$CuO$_4$, with $x = 1/8$. Here, an order of magnitude drop in the inplane resistivity marks the transition into some form of vortex liquid, while the interplane resistivity continues to grow with decreasing temperature. While a relatively weak diamagnetic signal persists to a higher temperature $T_M$ (similar to what is observed in La$_{2-x}$Sr$_x$CuO$_4$, with $x \sim 1/8$), in the present case, a sharp increase in the diamagnetic contribution to the susceptibility occurs at the same temperature, $T'_M$, as the resistive anomaly. ($T'_M$ is also the point at which spin-stripe ordering occurs, as determined by neutron scattering.) $T_{2D}$ appears dramatically in this material as the point at which the in-plane resistivity becomes immeasurably small, although the interplane resistivity remains large, so that for a range of temperatures below $T_{2D}$, this material appears to be literally a 2D superconductor. Finally, $T_c$ is the point at which flux expulsion, in the sense of perfect diamagnetism, first appears. (There is an intermediate temperature, $T_{2D} > T_{3D} > T_c$, below which the interplane resistivity becomes immeasurably small.)

In the case of the superconductor La$_{2-x}$Ba$_x$CuO$_4$, we have presented a plausible theoretical scenario in which the existence of a spatially modulated superconducting phase, which we called a pair-density-wave (PDW), produces the observed anomalies [9, 10]. Because the PDW order oscillates in sign, in many geometries [10] the interplane Josephson coupling can cancel exactly, producing an emergent layer decoupling and a delicate, 2D superconducting state. Moreover, in contrast to a usual superconductor, the long-range phase coherence of this phase is, generically, disrupted by even very weak disorder, possibly leading to a superconducting glass phase. Only when a usual, uniform, component of the superconducting order parameter develops can 3D correlations begin to grow, leading to a superconducting phase with true long-range order. While there are many aspects of the data that warrant further study, it seems to us likely that the same circle of ideas can account for the anomalous magnetization data in the bismuth-based high-temperature superconductors.

The observation of a clear thermodynamic signature of local superconducting order in a substantial portion of the pseudogap regime is an important step forward for the field. However, the pseudogap is often observed to onset at a significantly higher temperature, $T^* > T_M$, as shown schematically in Fig. 1. It it is also clear, from a variety of experiments, that other types of ordered states contribute significantly to the changes that occur at $T \sim T^*$. For instance, the recent studies of Daou *et al.*[2] reveal the growth of a large, anisotropic component of the Nernst effect below $T^*$ that continues to grow through $T_M$, suggestive of an intrinsic electronic tendency to spontaneous breaking of the point-group symmetry—electron nematic order [11]. Sorting out the relations between the different types of order and order-parameter fluctuations remains a key open problem.

## About the Authors

### Steven A. Kivelson

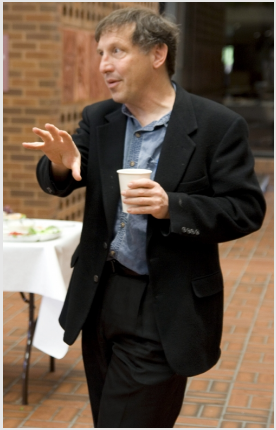

Steven Kivelson is a Professor of Physics at Stanford University. He graduated from Hill and Dale Nursery School, and some time later received his Ph.D. in physics from Harvard University. He is interested in the emergent properties of highly correlated fluids, both classical and quantum mechanical, and occasionally thinks about quenched disorder as well.

### Eduardo H. Fradkin

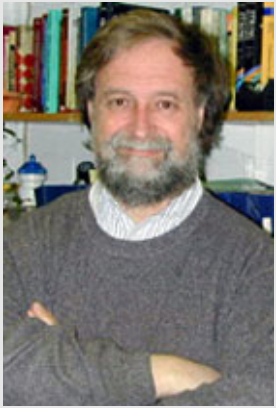

Eduardo Fradkin is a Professor of Physics at the University of Illinois at Urbana-Champaign. He received his Ph.D. in physics from Stanford University in 1979. He has worked on problems in condensed matter physics and gauge theory. He is interested in the physics of strongly correlated systems such as high $T_c$ superconductors and in topological phases of quantum condensed matter.